\documentclass[
reprint,
10pt,aps,
notitlepage, 
prb,
twocolumn,
showpacs,preprintnumbers,superscriptaddress,
amsmath,amssymb,floatfix]{revtex4-1}
\usepackage{graphicx}
\usepackage{dsfont}
\usepackage{dcolumn}
\usepackage{bm}
\usepackage{color}
\usepackage{xcolor}
\usepackage{url}
\usepackage{tabularx}
\usepackage{array}
\usepackage{booktabs}
\usepackage{comment}
\usepackage{hyperref}
\usepackage{footnote}
\usepackage{accents}
\usepackage{lipsum}
\usepackage[normalem]{ulem}
\DeclareMathOperator{\arccot}{arccot}

\newcommand{\fz}{Peter Gr\"unberg Institut and Institute for Advanced Simulation, Forschungszentrum J\"ulich and JARA, 52425 J\"ulich, Germany}

\begin{document}

\title{Magnetic skyrmions, chiral kinks and holomorphic functions}

\begin{abstract}

We present a novel approach to understanding the extraordinary diversity of magnetic skyrmion solutions. Our approach combines a new classification scheme with efficient analytical and numerical methods. We introduce the concept of chiral kinks to account for regions of disfavoured chirality in spin textures, and classify  two-dimensional  magnetic skyrmions in terms of closed domain walls carrying such chiral kinks. 
In particular, we show that the topological charge of magnetic skyrmions can be  expressed  in terms of the constituent  closed domain walls  and chiral kinks.
Guided by our classification scheme, we  propose a   method for  creating  hitherto unknown  magnetic skyrmions which involves  initial spin configurations formulated in terms of holomorphic functions  and subsequent numerical energy minimization.
We numerically   study the stability of the resulting magnetic skyrmions for a range of external fields and anisotropy parameters, and provide  quantitative estimates of the stability range for the whole variety of skyrmions with kinks. We show that the parameters limiting this range can be well described in terms of the relative energies of particular skyrmion solutions and isolated stripes with and without chiral kinks.

\end{abstract}

\author{Vladyslav~M.~Kuchkin}
\email{v.kuchkin@fz-juelich.de}
\affiliation{\fz}
\affiliation{Department of Physics, RWTH Aachen University, 52056 Aachen, Germany}

\author{Bruno~Barton-Singer}
\affiliation{Maxwell Institute for Mathematical Sciences and Department of Mathematics, Heriot-Watt University, Edinburgh, EH14 4AS, UK}

\author{Filipp~N.~Rybakov}
\affiliation{Department of Physics, KTH-Royal Institute of Technology, SE-10691 Stockholm,  Sweden}

\author{Stefan Bl\"ugel}
	\affiliation{\fz}
	
\author{Bernd J. Schroers}
\affiliation{Maxwell Institute for Mathematical Sciences and Department of Mathematics, Heriot-Watt University, Edinburgh, EH14 4AS, UK}

\author{Nikolai S. Kiselev}
	
	\affiliation{\fz}

\date{\today}

\maketitle 

\section{INTRODUCTION}

Chiral magnets are special  magnetic materials where the ground state is a homochiral spin spiral --  a helical or cycloidal modulation of the normalised magnetisation, $\textbf{n}(\textbf{r})\!=\!\mathbf{M}/|\mathbf{M}|$. 
The period of  modulations is determined by the ratio of the coupling constants of the Heisenberg exchange interaction and the chiral Dzyaloshinskii-Moriya interaction (DMI)~\cite{Dzyaloshinskii, Moriya},   modelled by a term  which we denote $w_\mathrm{D}(\textbf{n})$. 
The presence of the potential energy term, $U(\mathbf{n})$, in the 
Hamiltonian, $\mathcal{E}(\textbf{n})$, allows for the existence of magnetic solitons~\cite{Bogdanov_89} -- localized stable configurations possessing particle-like properties~\cite{Manton_04}.
Typical energy terms that contribute to $U(\textbf{n})$ are the interaction with the external magnetic field and the magnetocrystalline anisotropy. Note, the latter can also be thought of as an approximation of dipole-dipole interaction in ultrathin films~\cite{Gioia_James,Muratov_Slastikov}.

The solitons which arise in such materials can be classified by topological charges defined in terms of homotopy theory, and are therefore examples of topological solitons~\cite{Manton_04}.
In analogy with the topological solitons in the model of baryons proposed by T.\,H.\,R.~Skyrme~\cite{Skyrme},  it is  now common to refer to the topological solitons in chiral magnets as \textit{chiral magnetic skyrmions}\cite{Rossler_06}, or more simply as chiral skyrmions.

The standard  approach to finding  skyrmion solutions  is  based on the general variational principle, $\delta\mathcal{E}/\delta\textbf{n}(\mathbf{r})=0$, and  direct energy minimization with respect to all possible configurations of the corresponding field $\textbf{n}(\textbf{r})$. 
Due to the complexity of the problem the diversity of the solutions for chiral skyrmions was underestimated for a long time.
A variety of skyrmion solutions first were demonstrated by Bogdanov and Hubert in Ref.~\onlinecite{Bohdanov_99} where the authors studied so-called $k\pi$-skyrmions possessing two values of topological charge, $Q=0$ and $-1$ for even and odd $k$ respectively.
Numerical evidence for the existence of chiral skyrmions with arbitrary topological charge has been provided only recently in Refs.~\cite{Rybakov_19} and \cite{Foster_19}.
Although the skyrmions with complex morphology predicted in Ref.~\onlinecite{Rybakov_19} have not yet been observed experimentally in chiral magnets, the existence of similar textures has been proven by direct observation in liquid crystals~\cite{Foster_19}.

According to the Hobart-Derrick theorem, the stability of magnetic skyrmions requires the negativity of the integral of the  DMI energy term $w_\mathrm{D}(\textbf{n})$.
However, this requirement does not rule out local  variations in  the sign of the integrand, and so the DMI energy density of stable  skyrmion may be locally positive. In this paper we refer to regions where  
$w_\mathrm{D}(\textbf{n})<0$
as regions of favoured chirality and regions where 
$w_\mathrm{D}(\textbf{n})>0$
as regions of disfavoured chirality.  The emergence of regions of disfavoured chirality is natural for instance in the skyrmion lattice where axial symmetry of individual skyrmions is slightly distorted due to to inter-skyrmion interactions~\cite{McGrouther_16,Kovacs_17}.
On the other hand, the stability of an isolated skyrmion with locally disfavoured  chirality  is less obvious.
For instance, isolated $k\pi$-skyrmions in perpendicular external field have negative DMI energy density in the whole space.

It was recently shown  in Ref.~\onlinecite{Kuchkin_20}  that skyrmions  with locally  disfavoured  chirality remain stable in a finite range of the external magnetic field and anisotropy.
Moreover, stable skyrmions  with locally disfavoured  chirality  and arbitrary positive $Q$  are  discussed in Refs.~\onlinecite{Barton-Singer_20, Schroers_20}, where  exact analytical solutions are  studied at a particular point  of the phase diagram known as the Bogomol'nyi point. For our purposes it is important that the exact solutions can be expressed in terms of a single  complex function of one complex argument (the holomorphic function referred to in our title),  which may be chosen arbitrarily. 
The presence of such locally disfavoured  chirality significantly affects not only the stability of isolated skyrmions and their dynamics but also dramatically changes the character of inter-particle interactions from repulsive to attractive~\cite{Kuchkin_20}.
In the present paper, we introduce the concept of chiral kinks to account for regions of disfavoured chirality in magnetic skyrmions like the ones discussed  in Refs.~\cite{Kuchkin_20,Barton-Singer_20, Schroers_20}. We demonstrate the fruitfulness of this  concept  by using it in a  new method for generating magnetic  skyrmions with chiral kinks,  and  discuss the  stability  of the solutions thus obtained for a wide range of parameters.

Our concept of chiral kinks is a generalisation of what was called ``domain wall skyrmions" in the papers~\cite{Cheng_2019, Li_2020}. Both concepts describe a full rotation of the magnetisation vector along a domain wall in a two-dimensional spin texture. These were considered theoretically for infinite, straight domain walls, but we define chiral kinks of any winding number for domain walls of arbitrary geometry, and show that the topological charge of a spin texture can be expressed in terms of the number of kinks residing on domain walls in that spin texture.  We prefer the term chiral kink to domain wall skyrmions because, as we shall show, chiral kinks on domain walls provide a description of a spin texture which is different from and in a certain sense dual to its interpretation in terms of skyrmions.

The work is organized as follows. 
In Section II we define  the model and discuss our  numerical method for direct energy  minimization.
In Section III we present our approach to the  classification of magnetic skyrmions based on the observation that any configuration of a planar magnet defines a family  of domain walls, and that the angle of the magnetisation along a wall relative to the tangent direction of the wall supports topological excitations which we call chiral kinks.  
We show that the  topological charge of any magnetic skyrmions can be expressed in terms of number of kinks and domain walls, weighted with appropriate signs.
We also show that  our approach for soliton classification is consistent with the concept of so-called Bloch lines which is  well established in the theory of magnetic bubble domains. %
In Section IV we recall the general exact form  of magnetic skyrmions at the  Bogomol'nyi point,  and in Section V we use  families of such exact solutions  as initial states  at other points in the phase diagram for the direct energy minimization by means of numerical methods.
In Section V we analyse the energy dependence for different skyrmions as a  function of the external field and determine  the  range of external fields for which  the skyrmions with chiral kinks  are the lowest energy state for $Q\neq-1$.
Combining  analytical solutions and numerical analysis we discuss different classes of skyrmion solutions  and estimate  the range of their stability in Section VII.
In Section VIII we  provide estimates  for the characteristic size of chiral kinks, and in Section IX we discuss the range of optimal parameter where most of the presented solutions may coexist.
Our  final section  IX  contains a brief discussion and  concluding remarks.

\section{Model description}

The micromagnetic energy density functional for the two-dimensional (2D) chiral magnet is given by 
\begin{align}
{E}(\mathbf{n}) =\! \int\!\left(\mathcal{A}\left(\nabla\mathbf{n}\right)^{2}+\mathcal{D}\,w_\mathrm{D}(\mathbf{n})+ U(\textbf{n})\right)t\,\mathrm{d}x\mathrm{d}y, 
\label{eq:E_totm}
\end{align}
where $\mathcal{A}$ and $\mathcal{D}$ are micromagnetic constants of exchange interaction and DMI, respectively. 
The potential energy term $U(\textbf{n}) = U_\mathrm{a}(\textbf{n})+U_\mathrm{Z}(\textbf{n})$ includes uniaxial anisotropy, $U_\mathrm{a}(\textbf{n})\!=\!K(1-n_\mathrm{z}^2)$, and interaction with the external magnetic field, $U_\mathrm{Z}(\textbf{n})= M_\mathrm{s}B_\mathrm{ext}(1-n_\mathrm{z})$, applied perpendicularly to the plane along $\mathbf{e}_\mathrm{z}$. The magnetic texture assumed to be homogeneous along the film thickness, $t$.

The results presented in this work hold generally  for a broad class of magnetic crystals of different symmetry irrespective of whether the  DMI term favors Bloch or N\'eel type modulations. However, 
for definiteness,  we consider the  particular case of  $w_\mathrm{D}(\mathbf{n})\!=\!\mathbf{n}\!\cdot\!\nabla\!\times\!\mathbf{n}$ and $\mathcal{D}\!>\!0$, which favours right-handed Bloch-type modulations. 
This choice  of symmetry also allows us to illustrate the consistency of our approach with the concept of Bloch lines which is  well-established for magnetic bubble domain materials where Bloch type modulations are favored by dipole-dipole interactions.

The rescaling of spatial coordinates in units of equilibrium period of helical modulations, $L_\mathrm{D}=4\pi\mathcal{A}\mathcal{D}^{-1}$, and the value of external magnetic field in units of saturation field, $B_\mathrm{D}=\mathcal{D}^2/(2M_\mathrm{s}\mathcal{A})$,
allows one to write the functions~(\ref{eq:E_totm}) in dimensionless form
\begin{align}
\mathcal{E}(\mathbf{n}) = \int\left(\frac{\left(\nabla\mathbf{n}\right)^{2}}{2}+2\pi\,\mathbf{n}\cdot\nabla\times\mathbf{n}+U(n_\mathrm{z})\right)\mathrm{d}x\mathrm{d}y,
\label{eq:E_E0}
\end{align}
where the potential energy term is 
\begin{align}
U(n_\mathrm{z})=4\pi^{2}h(1-n_\mathrm{z}) + 4\pi^{2} u (1-n_\mathrm{z}^2).
\label{eq:U}
\end{align}
The dimensionless magnetic field and anisotropy are $h = B_\mathrm{ext}/B_\mathrm{D}$ and $u =K/(M_\mathrm{s} B_\mathrm{D})$, respectively. The energy is given in units of the energy of saturated state, $E_0\!=\!2\mathcal{A}\,t$.
The magnetization vector $\textbf{n}(\textbf{r})$ can be parameterized by spherical angles $\Theta(\textbf{r})$ and $\Phi(\textbf{r})$ as $\textbf{n}=\left(\sin\Theta \cos\Phi,\sin\Theta \sin\Phi, \cos\Theta\right)$.  

To find a stable solution representing a local or global minimum for the functional (\ref{eq:E_E0}), we use a
nonlinear conjugate gradient (NCG) method  with a finite-difference discretization scheme of the fourth order defined on a regular square grid with periodical boundary conditions~\cite{Rybakov_19}.
To achieve high accuracy in the estimation of the energies and stability of the solutions, we use large simulated domains with the size $\sim\!5 L_\mathrm{D}\!\times\!5L_\mathrm{D}$ or even higher, when necessary for a large size skyrmions.
The mesh density $\Delta l$ defined as a number of the mesh nodes per $L_\mathrm{D}$ can be controlled by the ratio $\mathcal{A}/\mathcal{D}$.
The typical values of $\Delta l$ are 64, 128, or 256 depending on the purposes and particular type of the solution and always specified in figure captions and in the main text.
We use various approaches to construct the initial spin-configuration followed by energy minimization with the GPU accelerated version of the NCG method implemented for the NVIDIA CUDA architecture (for details see Refs.~\cite{ Rybakov_19,Rybakov_15}).

\section{Domain walls and chiral kinks}

In the study of skyrmions - both in Skyrme's original  nuclear theory  and in condensed matter physics - it is customary to interpret  general field  configurations in terms of  constituent particle-like solitons, and to think of the  topological degree as counting the number of such solitons.  
However, one of the key messages of this paper is that configurations of chiral magnets in the plane which minimise the energy \eqref{eq:E_E0} are most naturally interpreted in terms of domain walls which may
carry  defects which we call chiral kinks.   The latter 
effectively represent  pairs of Bloch lines  of equal sign.  In this section we explain how the topological degree of an arbitrary  configuration can be expressed as a sum of winding numbers associated to domain  walls (which, in our two-dimensional setting, are of course one-dimensional) and obtain a relation between the winding number and the number of kinks on a domain wall.

\begin{figure}
\centering
\includegraphics[width=8cm]{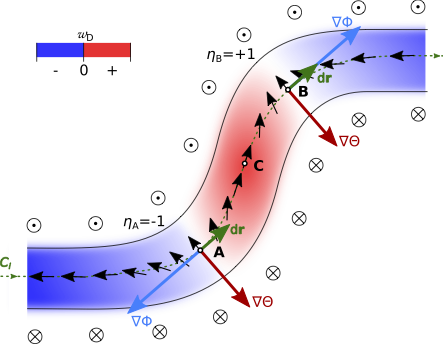}
\caption{\small 
Schematic representation of a section of the domain wall.  The black arrows are magnetization vector $\textbf{n}$ along the oriented contour $C_i$ (green dotted line) which divides the plane between the positive domain, $n_\mathrm{z}>0$ (on top) and negative domain $n_\mathrm{z}<0$ (at the bottom). The orientation of the contour 
$\mathrm{d}\mathbf{r}$ is chosen such that the positive domain is on the left and the negative domain is on the right. The red-white-blue color code represent variation of the chiral energy density which is vanishing at the position of Bloch lines marked as A and B. 
The handedness index of the Bloch lines is defined as $\eta = \mathrm{sign}(\nabla\Phi\cdot \mathrm{d}\mathbf{r})$. Along the path between A and B the vector $\nabla\Phi$ change the direction with respect to $\mathrm{d}\mathbf{r}$ and vanishes at point C while $w_\mathrm{D}$ at this point reaches maximal value.
%
}
\label{BW}
\end{figure}

Given a configuration $\mathbf{n}(\mathbf{r})$ in the plane, one can divide the plane into positive and negative domains according to whether $\mathbf{n}(\mathbf{r})$ takes  values in the upper ($n_\mathrm{z}>0$) or lower ($n_\mathrm{z}<0$) hemisphere at $\mathbf{r}$. These domains are separated by a region where $\mathbf{n}(\mathbf{r})$  takes values on the equator or, equivalently, $\Theta(\mathbf{r})= \frac \pi 2$ [Fig.~\ref{BW}]. For smooth configurations  and assuming maximal rank of the differential $\mathrm{d}\mathbf{n}$,  this region is a 1-dimensional submanifold of the plane, and therefore a countable union of simple (i.e. non-selfintersecting) curves  which are either closed or emerge from  and tend to infinity. 
In the following, we will refer to these curves or contour lines  as domain walls.
We should stress that the precise value of $\Theta$ on the contour does not matter for our purposes as long as $0\!<\!\Theta\!<\!\pi$. In particular, if the differential $\mathrm{d} \mathbf{n}$ happens to be degenerate when $\Theta=\frac \pi 2 $,   Sard's theorem assures that we can choose a value nearby where it is non-degenerate. 

We index the domain walls and corresponding contours by a countable index set $I$ and orient each contour so that it has a positive domain on the left and a negative domain on the right as illustrated in [Fig.~\ref{BW}].

Observing that the azimuthal angle $\Phi(\mathbf{r})$ is well-defined on a domain wall, we can therefore assign the winding numbers
\begin{align}
w(C_i) =\frac {1}{2\pi} \int_{C_i} \nabla \Phi \cdot \mathrm{d}\mathbf{r}, \qquad i\in I,
\label{WN}
\end{align}
to each domain wall $C_i $, with the direction of integration determined by the orientation of the  domain wall.  These numbers may be non-integer or even infinite for curves going off to infinity, however they are necessarily integers for closed curves. 

Generally, the degree of a configuration $\mathbf{n}(\mathbf{r})$, which can be defined analytically as 
\begin{equation}
Q(\mathbf{n}) = \frac{1}{4\pi}\int\!  \mathbf{n}\cdot \left(\partial_\mathrm{x}\mathbf{n}\times\partial_\mathrm{y}\mathbf{n}\right) \,\mathrm{d}x\mathrm{d}y,
\label{Qint}
\end{equation}
can  then  be expressed in terms of the winding numbers of the  domain walls as the sum
\begin{align}
   Q(\mathbf{n})= \sum_{i \in I}  w(C_i). 
   \label{QBloch}
\end{align} 
Note that the topological charge, $Q$, in \eqref{Qint} and \eqref{QBloch} may be infinite or ill-defined in general. However, when finite, the expressions \eqref{Qint} and \eqref{QBloch} agree.
The proof  of this result is provided in Appendix \ref{DWWN}.

\begin{figure*}[ht]
\centering
\includegraphics[width=15cm]{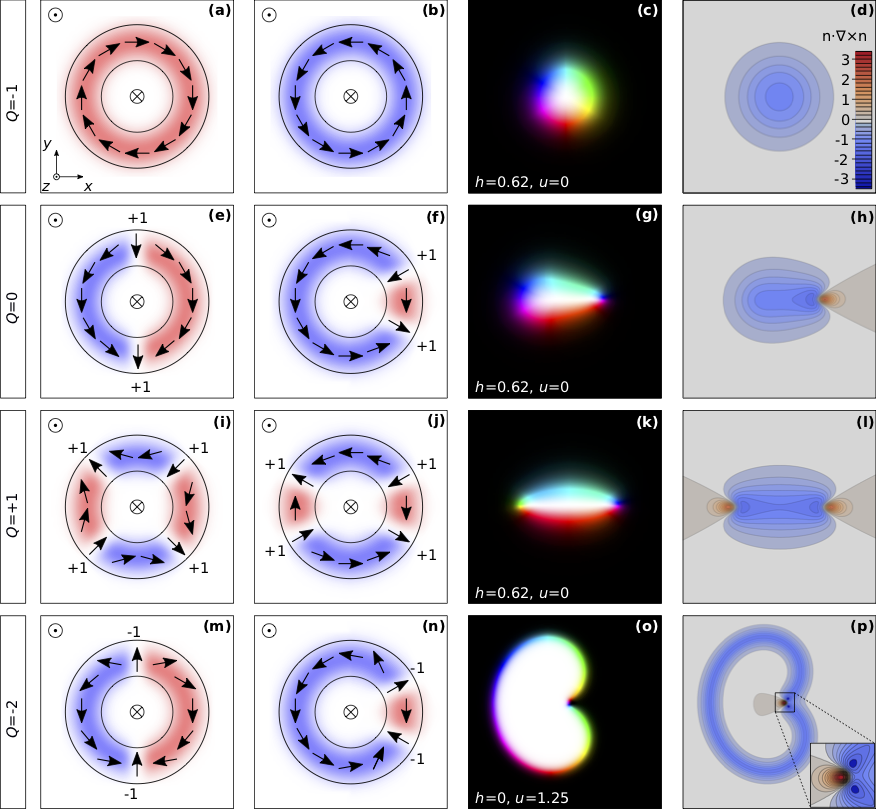}
\caption{~\small Magnetic skyrmions with chiral kinks in 2D magnet with DMI supporting right-handed Bloch-type modulations. Each row illustrates the skyrmions of a particular topological charge, as indicated on the left.
The first and the second columns illustrate schematically the orientation of the spins around the bubble domain core or skyrmion for $D\!\rightarrow\! 0$ and $D\!\neq\!0$, respectively.
The bluish and reddish regions in the schematic images indicate the negative and positive energy density of DMI respectively. 
The third column represents the spin texture of the corresponding skyrmion obtained by direct energy minimization.
All images are given in the same scale. 
We used a standard color code: black and white denote up and down spins respectively while red-green-blue reflect azimuthal angle, $\Phi$, with respect to $x$-axis.  The values of the external field, $h$, and uniaxial anisotropy, $u$, are indicated at the bottom of each image. 
The rightmost column contains the contour plots illustrating the distribution of scalar quantity $\mathbf{n}\!\cdot\!\nabla\!\times\!\mathbf{n}$ corresponding to the equilibrium spin-texture shown in the third column.
This scalar field up to a positively defined constant corresponds to the energy density of DMI.
}
\label{fig:1}
\end{figure*}

The winding number \eqref{WN} counts the winding of the magnetization vector $\mathbf{n}$ along a domain wall. The variation of $\mathbf{n}$  relative to the tangent direction of the domain wall may lead to a variation of the chirality of the domain wall.
For instance,  domain walls with alternate chirality are known to appear in ferromagnetic films with perpendicular anisotropy and strong dipole-dipole interactions favoring Bloch type modulations within the domain walls.
The transition regions with N\'eel like modulations which separate the regions of the Bloch domain wall with opposite chirality are known as Bloch lines~\cite{Malozemoff_79}. 
In this case, the position of the Bloch line on the domain wall is defined as the location where $\mathbf{n}$ is orthogonal to the tangent direction of the domain wall. 
Figure.~\ref{BW} illustrates a segment of the domain wall containing two Bloch lines at the points marked as A and B.
Note that  the Bloch lines in Fig.~\ref{BW} have  opposite handedness index, $\eta$, which is defined as a sign of the integrand in \eqref{WN}.
In the case of a chiral magnet, the position of the Bloch line can  be defined as the point which separates the regions of favoured chirality ($w_\mathrm{D}<0$) from the region of disfavoured chirality ($w_\mathrm{D}>0$). 
This definition holds generally,  for any type of DMI term.

In the case of closed domain walls, Bloch lines always appear in pairs otherwise the continuity of $\Phi(\mathbf{r})$ along the wall is broken.
Pairs of Bloch lines with opposite signs for $\eta$, called unwind pairs, always annihilate, while the pairs of the same sign for $\eta$  can be stable~\cite{Malozemoff_79}.
Such behaviour of Bloch lines can  easily be explained by means of topological arguments.
For instance, according to \eqref{WN} the topological charge of the texture depicted in Fig.~\ref{BW} is zero, essentially  because the intergrand in \eqref{WN} changes sign at point C.
Such a pair of Bloch lines represents an  unstable configuration which we have included for illustrative purposes.
On the other hand, for a pair of Bloch lines of the same handedness index, $\eta$, the  contribution to the topological charge is  $\pm1$ because the angle of $\mathbf{n}$ relative to the tangent direction of the wall  makes a $2\pi$ twist along the wall  in this case while  the sign of this contribution depends on the value of   $\eta$ of the Bloch lines or, in other words, on the direction of the twist.
Examples for stable pairs of Bloch lines on the closed domain wall are shown in Fig.~\ref{fig:1}.
In the absence of the DMI term,  the Bloch lines tend to be equidistantly distributed along the closed domain wall [Figs.~\ref{fig:1} (e), (i)  and (m)]. 
When the DMI term is nonzero, the system tends to reduce the distance between Bloch lines to minimize the area with disfavoured chirality and extend the area with  favoured chirality [Figs.~\ref{fig:1}(f), (j)  and (n)].
Thus, in chiral magnets, Bloch lines of the same sign `embrace' the region of disfavoured chirality and tend to form coupled states. In the following, we will refer to such coupled Bloch lines in chiral magnets as \textit{chiral kinks}.

\begin{figure*}[ht]
\centering
\includegraphics[width=15cm]{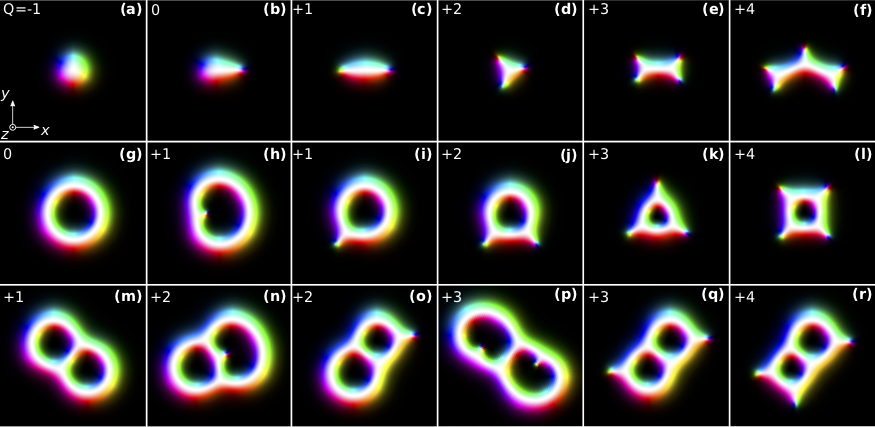}
\caption{~\small
Example of skyrmions with different topological charges (see the index in the left top corner) and the different number of chiral kinks.
The images (a), (g), and (m) in  the first column shows the host skyrmions with $Q=-1$, $0$, and $+1$.
Other images in each row show the skyrmions with a different number of kinks at different places of the host spin-texture.
The solutions are obtained by direct energy minimization on the domain with the size of $L_\mathrm{x}=L_\mathrm{y}=10L_\mathrm{D}$ and the mesh density $\Delta l = 64 $.
For all cases the anisotropy, $u=0$.
For skyrmion in (d) the external field $h=0.61$, for skyrmions in (a)-(c) and (e)-(f) $h=0.62$, and for skyrmions in (g)-(r) $h=0.64$.
}
\label{zoo1}
\end{figure*}

The topological charge of  spin textures which only comprise closed domain walls  can be expressed as a sum of  chiral kink numbers on the domain walls,  and this  will be useful for the following discussion.  Closed domain walls can be distinguished according to whether their orientation, as defined above, agrees or disagrees with their usual geometrical orientation (where the inside is always on the left), and we call the such walls positive in the first case and negative ins the second.
For instance, the domain walls in Fig.~\ref{fig:1} are all negative. 

 In Appendix \ref{DWWN},  we  define the chiral  kink number $ N_{\text{\tiny kink}}(C)$  and show that is related to the winding number $w(C)$ of $\Phi$ around $C$ by 
\begin{align}
N_{\text{\tiny kink}}(C) = w(C) -\iota(C),
\label{kinknumber}
\end{align}
where $\iota(C)$ is the  geometrical winding number of the wall. By Hopf's Umlaufsatz, $\iota(C)=\pm1$ for simple, closed walls,  with the sign equal to the sign  of the wall in  our convention. 
Hence we have $N_{\text{\tiny kink}}(C) = w(C) -1$ when such walls are positive  and $N_{\text{\tiny kink}}(C) = w(C) +1$ when they are negative.
Note that the kink number  on a domain wall is the sum of  kinks  on that wall weighted with their handedness index $\eta$, and therefore can be positive and negative.
For a  configuration $\mathbf{n}$ where all domain walls are closed, we split the index set $I$ into a disjoint union of $I^+$ and $I^-$ so that $C_i$ is positive (negative) for $i\in I^+$ ($i\in I^-$). 
Then we  can also  express the topological charge for the whole spin texture in terms of the  number of kinks hosted by the negative and positive domain walls as
\begin{align}
Q(\mathbf{n})= 
\sum_{i \in I^+} ( N_{\text{\tiny kink}}(C_i)+1) +\sum_{i \in I^-} (N_{\text{\tiny kink}}(C_i)-1).
   \label{Qkink}
\end{align} 
This formula is the topological basis for our interpretation of topologically non-trivial configurations of chiral magnets in terms of chiral kinks on domain walls in the rest of the paper.

In contrast to bubble domains, which usually remain axially symmetric in the presence of Bloch lines [Figs.~\ref{fig:1}(e), (i)  and (m)], the appearance of CKs in the structure of the chiral skyrmions leads to the violation of the axial symmetry of its spin texture.
One of the reasons for such discrepancies between bubble domains and chiral skyrmions is that the diameter of the bubble domains is usually a few times large than the width of the domain wall.
Moreover, the dipole-dipole interaction responsible for the stability of bubble domains but not essential for chiral skyrmions is a long-range interaction. By contrast, the DMI is a short-range local interaction.

Equilibrium textures representing magnetic skyrmions with different numbers and signs of CKs and the corresponding topological charges $Q$  are shown in the third column of Fig.~\ref{fig:1}. 
These magnetic textures were obtained by direct energy minimization of functional (\ref{eq:E_E0}) at $h$ and $u$ values indicated in the figures. 
The anti-skyrmion with $Q=+1$ in Fig.~\ref{fig:1} (k) was previously presented in Ref.~\onlinecite{Kuchkin_20} where we discussed the stability of chiral skyrmions in a tilted magnetic field.
We note that, for the spin textures with CKs, all derivatives of type  $\partial \mathbf{n}/\partial r_i$ ($i=x,y$)  which enter the exchange term and DMI in (\ref{eq:E_E0})  turn out to be  bounded at all points of the simulated domain.

The rightmost column in Fig.~\ref{fig:1} illustrates the distribution of the DMI energy density around the skyrmions. The key feature of the skyrmions which contain the CKs is the presence of regions with disfavoured chirality. 
As seen in the contour plots for the skyrmions with CK [Figs.~\ref{fig:1} (h), (l), and (p)] the emergence of red  areas with disfavoured chirality is accompanied by the development of  dark blue  regions of strongly favoured chirality, i.e. regions where $w_\mathrm{D}$ is more negative than for the skyrmion without CKs [Fig.~\ref{fig:1} (d)].
This  reflects a sophisticated balance between different energy terms which is responsible for the stability of such solutions.
One important consequence of the presence of the areas with  disfavoured chirality  is   the possibility of attractive skyrmion interactions. 
It was observed in \cite{Kuchkin_20}, for instance,  that  in a tilted magnetic field ordinary solutions for $\pi$-skyrmions without CKs lose their axial symmetry and at the same time develop areas of disfavoured chirality,   and that this in turn may lead  to the formation of a stable bound  pair of interacting skyrmions.
Note also that a careful study of the asymptotic behavior of the solutions for skyrmions with positive CKs [see Figs.~\ref{fig:1}(k) and \ref{fig:1}(l)] shows that regions of disfavoured chirality may extend to infinity~\cite{Kuchkin_20}.
On the other hand in the case of negative CK [Figs.~\ref{fig:1} (o) and (p)] the region of disfavoured chirality  is screened by a  region with favoured chirality.
Because of this screening effect, the skyrmions depicted in Fig.~\ref{fig:1}(o) are mutually repulsive, just like the axially symmetric $\pi$-skyrmions without kinks [Fig.~\ref{fig:1}(c)].

Before turning to the systematic investigation of a wide diversity of skyrmions with CKs in the next sections we look at Fig.~\ref{zoo1}  for a further illustration of the concept yielding the definition of topological charge in \eqref{QBloch} and \eqref{Qkink}.

The first column in Fig.~\ref{zoo1} shows the skyrmions without kinks representing the host spin-textures for various skyrmions with CKs shown in the other  columns. The Bloch walls lie in the coloured regions and, with our convention, should be traversed so that the black region lies on the left and the white region on the right. Comparing with the usual geometrical orientation, one finds, for example, that Fig.~\ref{zoo1}~(m) show one negative and two positive  walls. There are no kinks in this configuration, so $Q=2$ in agreement with \eqref{Qkink}. All the chiral kinks shown in  Fig.~\ref{zoo1} are positive, i.e. the handedness index, $\eta$, is positive when traversing the domain wall according to our convention. Therefore,  again by \eqref{Qkink}, the charge $Q$  equals the number of chiral kinks plus the number of positive Bloch walls and minus the number of negative Bloch walls.

It is obvious from the pictures that the presence of chiral kinks deforms the Bloch wall on which they reside. Moreover, the positive chiral kinks shown in Fig.~\ref{zoo1} produce an {\em inward} dent on positive walls and an  {\em outward} dent on negative walls. As we show in a separate study~\cite{letter}, his can be understood in terms of a simple effective theory for the kink field.
Some of the observations we made  about the  skyrmions  with single domain walls in Fig.~\ref{fig:1} generalise to the more intricate skyrmions in Fig.~\ref{zoo1}(g)-(r),  which are  composed of a several closed domain walls with a  varying numbers of kinks.
The skyrmions shown in Fig.~\ref{zoo1} can be divided into those which have CKs on their outer walls and those that do not. We have found that this division affects the interaction of pairs of such skyrmions.  The skyrmions with CKs on their outer Bloch walls have regions of disfavoured chirality stretching to spatial infinity. 
As for the antiskyrmion in Fig.~\ref{fig:1}~(l), this leads to such skyrmions being mutually attractive. By contrast, skyrmions with no CKs on their outer shells show the screening effect  discussed above for the  skyrmion with $Q=-2$ in  [ Fig.~\ref{fig:1}(o)-(p)]. In a perpendicular external field, such skyrmions always appear to repel each other.

\section{Skyrmions at the  Bogomol'nyi point}
One of remarkable properties of the chiral magnet model (\ref{eq:E_E0}) is that for parameters $h\!=\!1$ and $u\!=\!-0.5$ -- the Bogomol'nyi point, the model becomes exactly solvable~\cite{Barton-Singer_20}.
In this section, we review the key steps of Ref.\onlinecite{Barton-Singer_20} and Ref.\onlinecite{Schroers_20} and introduce a few classes of analytical solutions for skyrmions which are used below as the initial state for direct energy minimization with numerical methods for $h$ and $u$ outside the Bogomol'nyi point.

As observed in Ref.\onlinecite{Schroers_20}, the energy functional for chiral magnets with general type of DMI and potential term can be reinterpreted
as that of a gauged sigma model with a non-abelian gauge field determined by the DMI term. For instance, for the DMI term, $w_\mathrm{D}(\mathbf{n})\!=\!\mathbf{n}\!\cdot\!\nabla\!\times\!\mathbf{n}$, considered here, the required gauge field is  \(\mathbf{A}_\mathrm{x(y)}= -2\pi \mathbf{e}_\mathrm{x(y)}\). The field strength of this gauge field is  \(\mathbf{F}_\mathrm{xy} = \partial_\mathrm{x} \mathbf{A}_\mathrm{y} - \partial_\mathrm{y}\mathbf{A}_\mathrm{x} + \mathbf{A}_\mathrm{x}\times\mathbf{A}_\mathrm{y}\), so in our case, \(\mathbf{F}_\mathrm{xy} = \mathbf{A}_\mathrm{x}\times\mathbf{A}_\mathrm{y} = 4\pi^2 \mathbf{e}_\mathrm{z}\). This gauge field encodes the twisting of spins favoured by the balance of DMI and symmetric exchange, which is directly related to the microscopic quantity \(\mathcal{D}\mathcal{A}^{-1}\). The energy  of chiral magnets can be written in terms of the  covariant derivative \(D_\mathrm{x(y)} \mathbf{n} = \partial_\mathrm{x(y)}\mathbf{n} + \mathbf{A}_\mathrm{x(y)} \times \mathbf{n}\), 
the field strength and potential energy terms. It takes a particularly simple form at a special point in the phase diagram called the Bogomol'nyi point. For the energy  (\ref{eq:E_E0})  defining our model, the Bogomol'nyi point is at  $h\!=\!1$ and $u\!=\!-0.5$, where  the potential energy term (\ref{eq:U}) takes form
\begin{align}
U(n_\mathrm{z}) = 2\pi^{2}(1\!-\!n_\mathrm{z})^2, 
\end{align}
and the energy can be written as 
\begin{align}
E_\mathrm{c}(\mathbf{n})= \int\left(\frac{1}{2}\left(D_\mathrm{x} \mathbf{n}\right)^2+\frac{1}{2}\left(D_\mathrm{y} \mathbf{n} \right)^{2} -\mathbf{n}\cdot \mathbf{F}_\mathrm{xy}\right)\mathrm{d}x\mathrm{d}y.
\label{eq:Ec}
\end{align}
Applying the usual Bogomol'nyi trick, this can be expressed as 
\begin{gather}
E_\mathrm{c}(\mathbf{n}) = \frac{1}{2}\int\left(D_\mathrm{x} \mathbf{n} + 
\mathbf{n}\times D_\mathrm{y} \mathbf{n}\right)^{2}\mathrm{d}x\mathrm{d}y \nonumber \\ +4\pi Q(\mathbf{n})+\Omega(\mathbf{n}),
\end{gather}
where
\begin{gather}
\Omega(\mathbf{n})= \int\left(\partial_\mathrm{y} (\mathbf{n}\cdot \mathbf{A}_\mathrm{x}) - \partial_\mathrm{x}(\mathbf{n}\cdot \mathbf{A}_\mathrm{y}) \right)\mathrm{d}x\mathrm{d}y
\nonumber \\
=2\pi \int\left(\partial_\mathrm{x} n_\mathrm{y}- \partial_\mathrm{y} n_\mathrm{x} \right)\mathrm{d}x\mathrm{d}y,
\end{gather}
and $Q(\mathbf{n})$ is defind in (\ref{Qint}).

Since the term $\Omega(\mathbf{n})$ does not contribute to the Euler-Lagrange equations (see Appendix) it can be subtracted from the $E_\mathrm{c}(\mathbf{n})$ 
\begin{gather}
\tilde{E}_\mathrm{c}(\mathbf{n})=E_\mathrm{c}(\mathbf{n}) - \Omega(\mathbf{n}) \nonumber \\
=\frac{1}{2}\int\left(D_\mathrm{x} \mathbf{n} + \mathbf{n}\times D_\mathrm{y} \mathbf{n}\right)^{2}\mathrm{d}x\mathrm{d}y +4\pi Q(\mathbf{n}).
\label{tildeEc}
\end{gather}
The subtraction of $\Omega(\mathbf{n})$ ensures that the configuration $\mathbf{n}$ representing a solution of the Euler-Lagrange equations is the stationary point of the energy (for details see Appendix).

As follows from (\ref{tildeEc}) the energies of the soliton solutions are bounded from below
\begin{align}
\tilde{E}_\mathrm{c}(\mathbf{n})\geq 4\pi Q,
\label{E_bound}
\end{align}
which is known as the Bogomol'nyi bound. The inequality (\ref{E_bound}) becomes equality when the first term in (\ref{tildeEc}) vanishes:
\begin{equation}
D_\mathrm{x}\mathbf{n}+\mathbf{n}\times D_\mathrm{y}\mathbf{n}=0. 
\label{eq:BE}
\end{equation}
As shown in Ref.\onlinecite{Barton-Singer_20},   if a  configuration satisfies this  `Bogomol'nyi equation', then it is also a solution to the full Euler-Lagrange equations for the functional (\ref{eq:Ec}).

In the following, instead of vectors $\mathbf{n}=(n_\mathrm{x},n_\mathrm{y},n_\mathrm{z})$ and spatial coordinates $\mathbf{r}=(x, y)$ given in Cartesian coordinates, it is convenient to use the stereographic  projection of $\mathbf{n}$ into the complex plane and complex coordinates for  $\mathbf{r}$  according to
\begin{align}
w=\frac{n_\mathrm{x} + i n_\mathrm{y}}{1-n_\mathrm{z}}, \quad \zeta = x + i y.
\label{compc}
\end{align}
In  complex coordinates the Bogomol'nyi equation (\ref{eq:BE}) has the following form:
\begin{align}
\frac{\partial w}{\partial \bar{\zeta}}=i\pi w^2 \Leftrightarrow \frac{\partial (1/w)}{\partial \bar{\zeta}}=-i\pi. 
\end{align}
The general solution for above equation is:
\begin{align}
\frac{1}{w} = -i\pi \bar{\zeta} + f(\zeta),
\label{BG_solution}
\end{align}
where \(f\) is an arbitrary  holomorphic function  of the spatial coordinate $\zeta$.
The solution $w(\zeta)$ which satisfies the equation  (\ref{BG_solution})  can be represented in terms of Cartesian vectors  $\mathbf{n}(\mathbf{r})$ by inverting the maps \eqref{compc}:
\begin{align}
&\mathbf{n} = \frac{1}{1+|w|^2}\left(w+\bar{w},-i(w-\bar{w}),1-|w|^2\right),\\
&\mathbf{r} = \frac{1}{2}\left(\zeta+\bar{\zeta},-i(\zeta-\bar{\zeta})\right).
\label{n_mag}
\end{align}

\section{Skyrmions beyond the exact solvable model}

Now we consider some illustrative examples of  the functions $f(\zeta)$ which can be used as an ansatz (initial configuration) for the numerical minimization of the energy functional (\ref{eq:E_E0}).
First, we discuss a wide class of solutions   composed of functions  $f(\zeta)$ which are a ratio of two polynomials of order $m$ and $n$, respectively. The topological charge $Q$ of the corresponding configuration (\ref{n_mag}) is finite and can be computed as follows \cite{Degree}, \cite{Bleher}. If $m>n+1$, then $Q=m$. If $m<n+1$, $Q=n-1$. In the case where $m=n+1$, the behaviour depends on the behaviour of $f(\zeta)$ as $\zeta$ becomes large:
\begin{equation}
\label{BogSolnCharges}
Q= \begin{cases}
n+1, & \lim_{\zeta\to\infty}f > \pi, \\
n, & \lim_{\zeta\to\infty}f = \pi, \\
n-1, & \lim_{\zeta\to\infty}f < \pi, 
\end{cases}
\end{equation}

We can turn this around and ask for the number of degrees of freedom we have to construct a solution with a given $Q$. This is straightforward when $m \neq n+1$. In the case $m=n+1$, there are three possible general functions:
\begin{align}
f(\zeta) &= \frac{a_Q \zeta^{Q} + \ldots +a_0 }{b_{Q-1}\zeta^{Q-1}+\ldots + b_0}, \lvert \frac{a_Q}{b_{Q-1}} \rvert >\pi,  \label{fz_gen1}\\
f(\zeta) &= \pi e^{i\alpha}\frac{\zeta^{Q+1}+\ldots + a_0}{\zeta^Q+\ldots+b_0}, \label{fz_gen2}\\
f(\zeta) &= \frac{a_{Q+2} \zeta^{Q+2} + \ldots + a_0}{b_{Q+1}\zeta^{Q+1} + \ldots+ b_0} ,\lvert \frac{a_{Q+2}}{b_{Q+1}} \rvert <\pi.
\label{fz_gen3}
\end{align}
where we assume that the leading coefficients of each polynomial is non-zero, and that numerator and denominator have no common factors.  

\begin{figure}
\centering
\includegraphics[width=8cm]{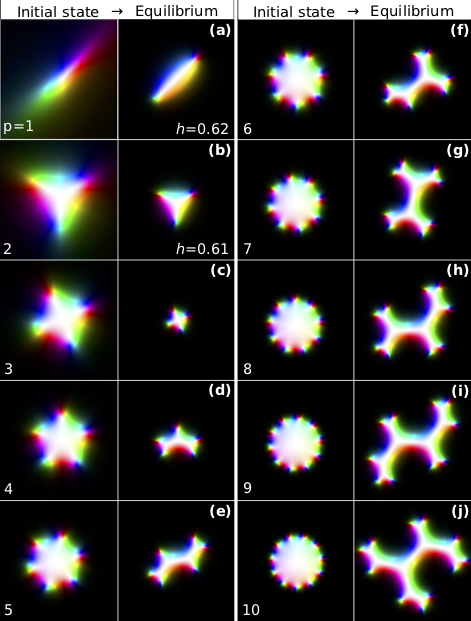}
\caption{\small The left image in each figure (a)-(e) is the initial magnetic textures which are described by (\ref{ini_ansatz}) where $f(\zeta)\!=\!2\zeta^p$ and the scaling parameters are $l_{2}=0.5$ and $l_{1}=1$.
The topological charge of these solutions is given by $Q\!=\!p$. The right image is the equilibrium magnetization corresponding to a local minimum obtained after numerical energy minimization. The parameters of the system: $h\!=\!0.62$ (f), $h\!=\!0.61$ (g) and $h\!=\!0.65$ for all others, $u=0$ in all cases. The size of simulated domain $8L_\mathrm{D} \times 8L_\mathrm{D}$ the mesh density $\Delta l\!=\!128$. 
}
\label{zoo2}
\end{figure}

We illustrate the general results with a  family of configurations which play an important role in discussion of instabilities later in this paper, namely the configurations determined by
\begin{equation}
f(\zeta) =\frac{a_2\zeta^2+ a_0}{\zeta}.  
\label{unstable}
\end{equation}
When $a_0=0$ and $|a_2| > \pi$ the corresponding field is that of  a single anti-skyrmion [Fig.~\ref{zoo2} (a)], so $Q=1$. Still keeping $a_0=0$ but taking the limit $|a_2|\rightarrow \pi$, the anti-skyrmion elongates and, when $|a_2|=\pi$,  becomes an infinite  isolated stripe (also called line defect \cite{Barton-Singer_20}). The direction of the isolated stripe depends on the phase of $a_2$. Taking $a_2=-i\pi$ for definiteness we obtain a  stripe parallel to the $y$-axis. The  magnetization across the stripe is conveniently expressed in terms of the angles $\Theta$ and $\Phi$ which are related to the complex field $v$ via
\begin{equation}
 \frac 1 w =\cot\frac\Theta 2 e^{-i\Phi}.   
\end{equation}
Thus the solution $1/w=-i\pi(\bar{\zeta}+ \zeta)=-2\pi i x$ can equivalently be written as 
\begin{equation}
    \Theta(x)= 2 \arccot(2\pi|x|),
 \;\; \Phi(x) = \begin{cases} \frac{\pi}{2} & \text{for} \;  x>0 \\ \frac{3\pi}{2} & \text{for} \;  x<0 \end{cases}.
 \label{bogowall}
\end{equation}
This reveals the geometrical shape of this solution:   the magnetisation vector $\mathbf{n}$ performs a complete rotation  in the plane orthogonal to the direction travel as one  traverses the $x$-axis, beginning and ending in the vacuum $\Theta=0$.

Finally switching on  the coefficient $a_0$ in \eqref{unstable} while keeping $a_2=-i\pi$, we obtain a configuration of degree $Q=1$ where an anti-skyrmion has been inserted at the origin,  and broken the  isolated stripe into two halves, both capped by half an anti-skyrmion [Fig.~\ref{fig:hu}(f)].  It turns out that the deformation of anti-skrymions into an isolated stripe and the rupture of the stripe  play an important role in our discussion of instabilities in Sect.~\ref{stabsect}.

\begin{figure}
\centering
\includegraphics[width=8cm]{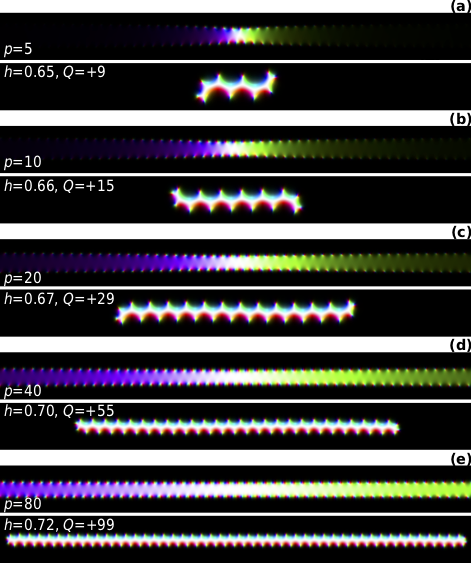}
\caption{~\small The top image in each figure (a)-(e) is the initial magnetic textures which are described by (\ref{ini_ansatz}) where  $f(\zeta)=0.025\sin(p\zeta)$ and the scaling parameters $2l_{2}=l_{1}=p/20$.
The bottom image is the equilibrium magnetization corresponding to a local minimum obtained after numerical energy minimization.
The parameters of the system: domain size $16L_\mathrm{D} \times 2L_\mathrm{D}$, mesh density $\Delta l=128$, $u=0$, the values of external magnetic field $h$ and topological charge $Q$ after full energy minimization are indicated in the images.
}
\label{zoo3}
\end{figure}

\begin{figure}
\centering
\includegraphics[width=8cm]{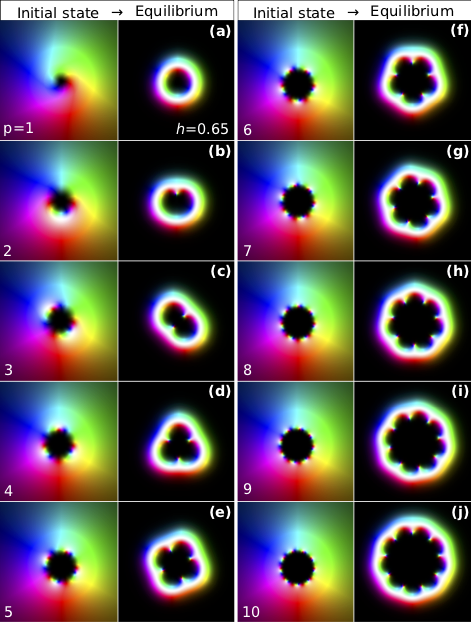}
\caption{
The left image in each figure (a)-(e) is the initial magnetic textures which are described by (\ref{ini_ansatz}) where $f(\zeta)=1/(2\zeta^{p})$ and the topological charge of the solution is $Q=p-1$.
The right image is the equilibrium magnetization corresponding to a local minimum obtained after numerical energy minimization. The parameters of the system: $h=0.65$, the size of simulated domain $8L_\mathrm{D} \times 8L_\mathrm{D}$ the mesh density $\Delta l\!=\!128$.
}
\label{zoo4}
\end{figure}

\begin{figure}
\centering
\includegraphics[width=8cm]{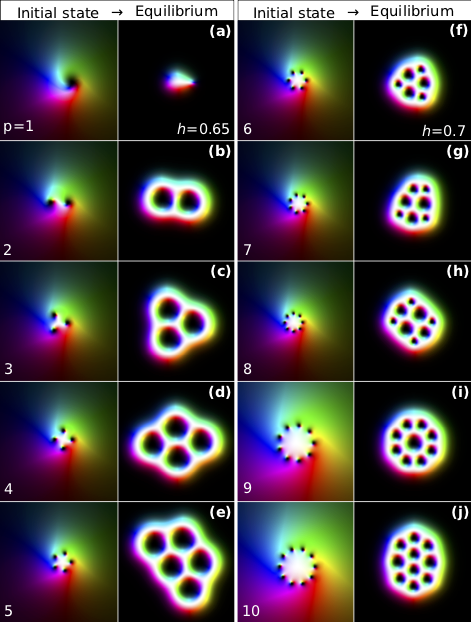}
\caption{
The left image in each figure (a)-(e) is the initial magnetic textures which are described by (\ref{ini_ansatz}) where  $f(\zeta)=\zeta^{p-1}/(\zeta^{p}-1)$, the topological charge of the solution $Q=p-1$. The right image is the equilibrium magnetization corresponding to a local minimum obtained after numerical energy minimization. The parameters of the system: the size of simulated domain $8L_\mathrm{D} \times 8L_\mathrm{D}$ the mesh density $\Delta l\!=\!128$, $u=0$.
For skyrmions in (a)-(e) the external field $h=0.65$, for (f)-(j) $h=0.7$.
}
\label{zoo5}
\end{figure}

We now turn to more general  functions of the form  (\ref{fz_gen1})-(\ref{fz_gen3}) and use them to produce  ansatz solutions away from the Bogomol'nyi point  in the form
\begin{align}
\frac{1}{w} = -\frac{i\bar{\zeta}}{l_1} + f\left(\frac{\zeta}{l_2}\right),
\label{ini_ansatz}
\end{align}
 where $l_{1}$ and $l_{2}$ are arbitrary scaling parameters chosen with respect to the size of simulated domain and the mesh density used in numerical scheme.  
The functions $f(\zeta)$ depending on their analytical properties, e.g. number of zeros and poles, provide the solutions for different classes of skyrmions Fig.~\ref{zoo2} -- Fig.~\ref{zoo5}.
 
First, we consider an example when function $f(\zeta)$ does not have poles, namely
 \begin{equation}
   f(\zeta)=2\zeta^{p},  
   \label{zeta^p}
 \end{equation}
where $p\in\mathbb{Z}^+$ -- positive integer numbers.
This ansatz describes axially symmetric skyrmions with $p+1$ positive CKs equidistantly distributed over the perimeter of the skyrmion, see the images for  the initial states in Fig.~\ref{zoo2}.
The corresponding equilibrium configurations obtained after numerical energy minimization are depicted in Fig.~\ref{zoo2} on the right.

There are a few interesting aspects related to these solutions.
In contrast to previously  studied skyrmion sacks that may coexist in a very wide range of parameters the morphologically similar solutions in Fig.~\ref{zoo2} are stable in different ranges.
For instance, in the case  $u=0$, the  anti-skyrmion obtained with  the ansatz (\ref{zeta^p}) with $p=1$ [Fig.~\ref{zoo2}(a)] and the  skyrmion with $p=2$ [Fig.~\ref{zoo2}(b)] are stable in non-overlapping ranges of the magnetic field. 
Moreover, with an increasing number of CKs (for $p>3$) the skyrmions of this type lose axial symmetry and tend to form  complex shapes  of   branching trees.
Because of that for $p>10$, the ansatz (\ref{zeta^p}) does not provide a good initial guess. To obtain such branched skyrmions one can consider generalized polynomials $f(\zeta)=\prod_{j=1}^{p}(\zeta - a_{j})$ with different roots $a_{j}$ in the complex plane. 

One can also consider trigonometric or exponential functions. They have an infinite number of zeros in the plane, but  finitely many in any finite  region. Such functions therefore provide a useful ansatz for obtaining skyrmions with large $Q$. 
Fig.~\ref{zoo3} illustrates stretched skyrmions of high $Q$ and large number of CKs obtained  from
\begin{equation}
f(\zeta)=\alpha \sin(p\zeta),\,\, p\in\mathbb{Z}^+,
\end{equation}
where $\alpha$ is arbitrary non-zero constant.
The existence of such skyrmions also suggests the stability of the CKs in isolated domain walls and stripes, which will be discussed in the following sections.

Another class of solutions can be obtained with the function $f(\zeta)$ that has poles but does not have zeros:
\begin{equation}
   f(\zeta)=1/(2\zeta^p), \, p\in\mathbb{Z}^+.
\end{equation} 
This class of solutions represents $2\pi$-skyrmions with $p-1$ positive  CKs (in our conventions) on its inner side. Fig.~\ref{zoo4} shows initial configurations  and corresponding equilibrium states obtained by direct energy minimization.
In contrast to (\ref{zeta^p}) this ansatz  gives a  satisfactory initial configuration for any value of $p$. 

Finally we consider the class of skyrmion `sacks' with high $Q$  discussed in Ref.~\onlinecite{Rybakov_19}.
Such solutions can be obtained in our scheme  by taking  functions $f(\zeta)$ which combine features of the previous two cases and which  have poles and zeros of order $p$ and $p-1$ respectively:
\begin{equation}
f(\zeta)=\zeta^{p-1}/(\zeta^{p}-1),\,\, p\in\mathbb{Z}^+.
\label{zeta4}
\end{equation}
The initial states defined by (\ref{zeta4}) and the  corresponding equilibrium configurations are shown in Fig.~\ref{zoo5}.
The topological charge of this class of solutions is  $Q=p-1$.
Among all skyrmions of this class only the skyrmion with $Q=0$ [Fig.~\ref{zoo5}(a)] has a CK while the  other skyrmions after energy minimization converge to the states free of CKs.

Like the ansatz (\ref{zeta^p}),  the ansatz (\ref{zeta4})  is only useful in a finite range of values for  $p$;  for $p>10$  it does not provide a good initial guess. 
To obtain configurations of this type but with higher $Q$,  one can use a more general class of   functions  $f(\zeta)$ given in~(\ref{fz_gen1}-\ref{fz_gen3}),  and set the explicit distribution of zeros $a_{j}$ and poles $b_{j}$ of $f(\zeta)$ via $f(\zeta)=\frac{1}{\zeta - b_{p}}\prod_{j=1}^{p-1}\frac{\zeta - a_{j}}{\zeta - b_{j}}$. With this approach one can get skyrmions with $Q$ higher than that for skyrmions shown in Fig.~\ref{zoo5}.

The approach presented in this section  for the construction of the initial states allows one to obtain a wide class of solutions, but  it also  has  limitations.
In particular, skyrmion sacks (skyrmion bags) representing a $2\pi$-skyrmion shell with skyrmion cores inside (the simplest example is a $3\pi$-skyrmion) can not be obtained using (\ref{ini_ansatz}). 
There is an explanation for this: at the Bogomol'nyi point, the $2\pi$-skyrmion solution has a zero mode corresponding to changing the size of the bag. In other words, the domain wall that forms the outside of the bag has no `tension'. Therefore if skyrmions are put inside it then the additional repulsive force will cause the bag to expand to infinity. One can also show that $k \pi$-skyrmions do not exist for $k\geq3$ at the Bogomol'nyi point, and that the solutions described by \eqref{BG_solution} do not include solitons with $Q\leq -2$.

To construct initial configurations for skyrmions which are not covered by Eqs.~(\ref{ini_ansatz}), 
we either used piecewise functions based on these equations or crafted a texture by means of interactive tools implemented in our software~\cite{Excalibur}(see also Supplemental Material in Ref.~\onlinecite{Rybakov_19}).

\section{Energies of skyrmions with chiral kinks}

\begin{figure}[h]
\centering
\includegraphics[width=8cm]{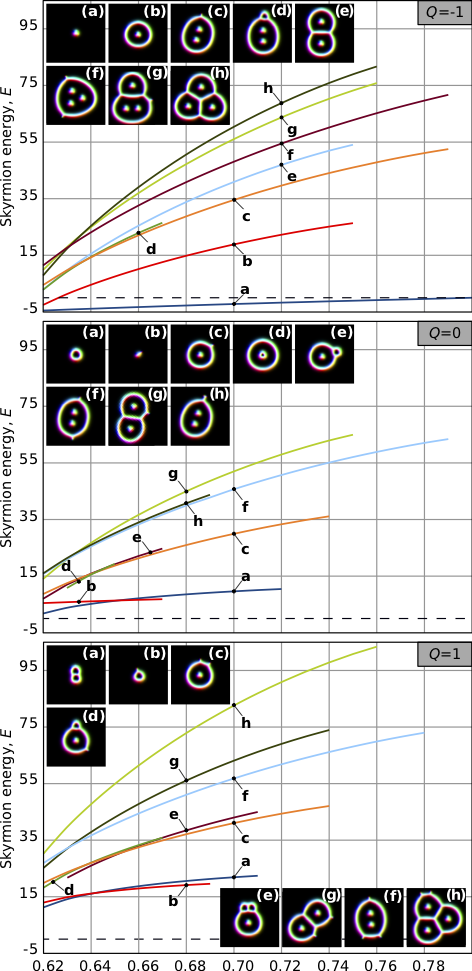}
\caption{~\small The self energy of skyrmions as function of the external field, $h$, for the case of zero anisotropy, $u=0$. Top, middle and bottom plots correspond to the skyrmions with $Q=-1$, $0$, and $+1$, as indicated in the right top corner.
The horizontal dashed line corresponds to $E=0$ -- the energy of field saturated state.
The parameters of the system:
the size of simulated domain $8L_\mathrm{D}\times 8L_\mathrm{D}$, the mesh density $\Delta l=64$.}
\label{EQ}
\end{figure}

Since the presence of CKs   indicates positive energy  contributions of the DMI  in some areas, one may guess that the skyrmions with CKs always have higher energies than  skyrmions without them. In this section, we show that this is not the case and  that solutions containing CKs may become energetically comparable   or even favorable relative to   skyrmions without CKs.
In particular, we have calculated the energy dependencies for skyrmions with and without CKs as a function of the external field and group them with respect to their topological charge, $Q=-1$, 0, and 1 [Fig.~\ref{EQ}].

For any $Q$, at low magnetic fields, $h \lesssim 0.62$, the shown skyrmions are  elliptically unstable. 
For high magnetic fields, the energy dependence curves in Fig.~\ref{EQ} end at the points which correspond to the collapse of the soliton. 
The only exception is the $\pi$-skyrmion with $Q=-1$. There is a strict mathematical proof that in the case of $u=0$ the $\pi$-skyrmion remains stable for any $h$ above the elliptical instability and represents the lowest energy soliton solution in this topological sector~\cite{Melcher}. 
Moreover,  for $h<0.82$ the energy of the $\pi$-skyrmion is lower than the energy of ferromagnetic state (horizontal dashed line). 
The latter indicates the well-known fact that the lattice of $\pi$-skyrmions becomes the ground state of the system~\cite{Bogdanov_1994JMMM}.

In the case of $Q=-1$ in Fig.~\ref{EQ} the  $\pi$-skyrmion is the universal minimizer -- the lowest energy state in the whole range of fields. The $3\pi$-skyrmion (b)  has the second lowest energy but collapses at $h>0.75$. 
At higher magnetic fields,  the solution with CK Fig.~\ref{EQ}~(c)  has the second lowest energy among  $Q=-1$ solutions. 
It is interesting that for $Q=-1$ the solutions without CKs [Fig.~\ref{EQ} (d), (e) and (h) for $Q=-1$] have higher energy and are stable in the narrower ranges of fields than the skyrmion with CK in Fig.~\ref{EQ}~(c).

In contrast to the situation in the $Q=-1$ sector, there are no universal minimizers in the  $Q=0$ and $Q=1$ sectors. 
In particular, for $Q=0$  and  a low  magnetic field,  the lowest energy state
is  the skyrmionium [Fig.~\ref{EQ}~(a)]. However, in the range of $0.65 \lesssim h \lesssim 0.67$ the lowest energy state is the skyrmion with one CK [Fig.~\ref{EQ}~(b)]. Note that the skyrmion with one CK in Fig.~\ref{EQ}~(b) is identical to that shown in Fig.~\ref{fig:1}~(g).
For $h>0.67$ this skyrmion is unstable and skyrmionium becomes the lowest energy state again.
Above the skyrmionium collapse field the lowest energy state corresponds to the skyrmion solution with CKs: the skyrmion in (c) in range of $0.71 \lesssim h \lesssim 0.74$ and the skyrmion in (f) for $0.74 \lesssim h \lesssim 0.79$, see Fig.~\ref{EQ} for $Q=0$.  

For the case of $Q=1$, the behavior of the solutions is very similar to that for $Q=0$.
The lowest energy state alternates between states (a) and (b) and when  the skyrmion in (a) collapses the lowest energy state corresponds to state (c) in the range of fields $0.71 \lesssim h \lesssim 0.74$ and skyrmion in (f) in the range of fields $0.74 \lesssim h \lesssim 0.78$, see Fig.~\ref{EQ} for $Q=1$. 
The skyrmions with $|Q|>1$ show nearly the same behaviour but because of the complex morphology and large diversity of the solutions with large $Q$ they are not discussed here. 

It is worth emphasizing that the range of the existence for skyrmions which are shown in Fig.~\ref{EQ} may increase when a more precise numerical scheme is used. 
Nevertheless, Fig.~\ref{EQ} clearly illustrates that the energies of skyrmions with and without CKs are comparable.
Thereby the solutions containing CKs cannot be excluded from consideration. 
Besides that one can conclude the more complicated morphology of skyrmions leads to the higher stability of the solutions at strong magnetic fields. 
Because the energy of skyrmions with CKs possess a few times higher energies than the energy of $\pi$-skyrmion such skyrmions cannot become the ground state of the system.

\section{Stability upper bound for skyrmions with chiral kinks}
\label{stabsect}

\begin{figure}[h]
\centering
\includegraphics[width=8cm]{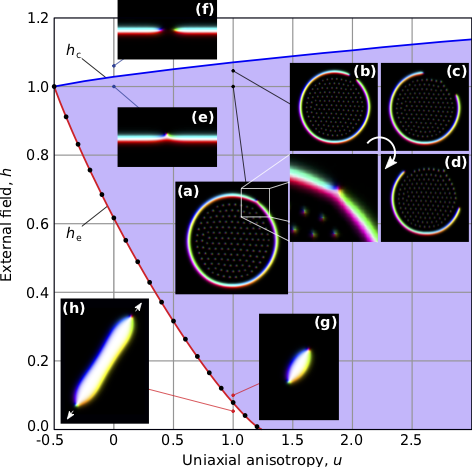}
\caption{~\small $h$-$u$ diagram of stability of magnetic skyrmions with CKs. 
(a) is an example of a large skyrmion with the negative CK in outer side of the shell at $h=1$, $u=1$.
(b)-(d) are the snapshots of the system at different time after a rupture of the shell with increasing the external field up to $h=1.05$.
The collapse field, $h_\mathrm{c}$ (blue solid line) corresponds to the field at which the isolated stripe with the negative CK is ruptured (e)-(f).
The elliptic instability field, $h_\mathrm{e}$ (red solid line) are defined by the criteria that the energy of isolated stripe in (\ref{DWenergy}) equals zero, $\mathcal{E}_\mathrm{IS}=0$.
The black dots lying very close to red line corresponds to numerically estimated elliptic instability of the anti-skyrmion (g).
For $h<h_\mathrm{e}$ the anti-skyrmion starts to elongate abruptly as indicated by wight arrows in (h).
}
\label{fig:hu}
\end{figure}

To estimate the range of stability for the skyrmions containing CKs, we consider the limiting case of a large skyrmion sack with $Q\ll-1$.
An example of such a skyrmion with one CK in the outer side of its shell is shown in Fig.~\ref{fig:hu}~(a).
We found that the skyrmions of this type are the most stable among all skyrmions possessing CKs.
At high magnetic fields, such skyrmions collapse via a rupture of the shell at the position where the CK was placed [Fig.~\ref{fig:hu}~(b)-(d)].
Such a rupture changes the number of kinks and walls, but maintains the overall charge $Q$ as the magnetization field is changing smoothly.
The ruptured shell of the skyrmion in Figs.~\ref{fig:hu}~(b)-(d) is identical to the elongated skyrmion depicted in Fig.~\ref{fig:1}~(k), which in this case will then collapse. 
The larger the size of the skyrmion, the higher the external field required for its collapse. 
Thereby, to estimate the stability of such solutions from the top we consider the limiting case of the isolated stripe with CK [Fig.~\ref{fig:hu}~(e)].
The blue line, $h_\mathrm{c}$, in Fig.~\ref{fig:hu} represents the collapse field for an isolated stripe with CK, estimated numerically with  high accuracy.

In order to provide estimates for the lower limiting field for skyrmions with CK we do the following. 
We consider the anti-skyrmion as a stripe capped by two chiral kinks. For the functional (\ref{eq:E_E0}) one can find the solution of the variational problem for an isolated stripe along the $y$-axis 
and thus calculate its energy per unit length, $\mathcal{E}_\mathrm{IS}(h,u)$. 
When $\mathcal{E}_\mathrm{IS}(h,u)<0$, the stripe has negative energy per unit length, so it will extend, and therefore so will an anti-skyrmion [Fig.~\ref{fig:hu}~(h)]. When $\mathcal{E}_\mathrm{IS}(h,u)>0$, the stripe has positive energy per unit length so will tend to be as short as possible, giving a stable shape for the anti-skyrmion [Fig.~\ref{fig:hu}~(g)].
Along the curve $\mathcal{E}_\mathrm{IS}(h,u)=0$ the extension of the anti-skyrmion into the stripe is a zero mode. 
The criterion $\mathcal{E}_\mathrm{IS}(h,u)=0$ gives a good estimation of the elliptic instability of the anti-skyrmion which can be understood by taking into account its elongated shape~ [Fig.~\ref{fig:hu}~(g)].
Higher $Q$ configurations can also be considered as stripes with several kinks attached, although this is a worse approximation. 

The solution for the isolated stripe can be written as follows~\cite{Meynell_14,Muller_16}:
\begin{align}
   \Theta(x)& = 2 \arccot\left(\frac{\sqrt{h}}{k} \lvert \sinh(2\pi k x)\rvert\right), \nonumber \\
 \Phi(x) & = \begin{cases} \frac{\pi}{2} & \text{for} \;  x>0 \\ \frac{3\pi}{2} & \text{for} \;  x<0 \end{cases}.
 \label{stripe_profile}
\end{align}
where $k=\sqrt{h+2u}$.
Taking the limit $k\to 0$ gives back the solution \eqref{bogowall} at the Bogomol'nyi point.

After integration over $x$, the energy per unit length of the isolated stripe is~\cite{Bogdanov_1994JMMM}
\begin{align}
\mathcal{E}_\mathrm{IS}=-4\pi^{2}+8\pi k+\frac{2\pi\sqrt{2}h}{\sqrt{u}}\log\frac{k+\sqrt{2u}}{k-\sqrt{2u}}.
\label{DWenergy}
\end{align}
As follows from (\ref{DWenergy}) the solution for the isolated stripe for  $h\geq0$ remains stable under the condition $h+2u\geq0$. 
The energy of the isolated stripe increases gradually with increasing $h$ and $u$.
The asymptotic behavior of the energy of the solution (\ref{DWenergy}) for $h$ and $u$ independently approaching infinity are
\begin{align}
\mathcal{E}_\mathrm{IS}(u)=16\pi^{2}u-4\sqrt{2}\pi^{3}\sqrt{u}+\mathcal{O}\left(\sqrt{u}\right),\ u\gg1,
\end{align}
for any fixed $h\geq0$,
and
\begin{align}
\mathcal{E}_\mathrm{IS}(h)=16\pi^{2}h-4\pi^{3}\sqrt{h}+\mathcal{O}\left(\sqrt{h}\right),\ h\gg1,
\end{align}
for fixed values of $u$.

The red solid line, $h_\mathrm{e}$, in Fig.~\ref{fig:hu} corresponds to $\mathcal{E}_\mathrm{IS}(h,u)=0$, while black dots are the elliptic instability field for anti-skyrmion estimated numerically [Figs.~\ref{fig:hu}~(g)-(h)].

The elliptic instability field, $h_\mathrm{e}$ and the collapse field $h_\mathrm{c}$ meet at Bogomol'nyi point ($h=1$, $u=-0.5$).
Thereby the stability of skyrmions with CKs is limited by the strong easy-plane anisotropy, $u=-0.5$.
On the other hand, there is no limiting value for strong easy-axis anisotropy above which the solutions for chiral skyrmions with CK would be unstable.
The latter statements can be proven as follows.

Let us consider an isolated straight stripe with one CK. The energy of such a solution can be bounded above by an ansatz where $\Theta(x)$ is as  in \eqref{stripe_profile}, and $\Phi$ depends on $y$ for $x>0$. Minimizing the energy within this ansatz gives:

\begin{align}
\Phi_{a}\left(x,y\right)= & \begin{cases}
\frac{3\pi}{2} & x<0,\\
4\arctan(e^{m y})+\frac{\pi}{2} & x>0,
\end{cases}
\end{align}
where the functional dependence of $m$ on $h$ and $u$ is given in Appendix \eqref{CKansatz}. The corresponding energy of the solution can be written as follows

\begin{align}
\mathcal{E}_\mathrm{IS+CK}=\mathcal{E}_\mathrm{IS}+\mathcal{E}_\mathrm{CK},
\end{align}
where self energy of isolated stripe $\mathcal{E}_\mathrm{IS}$ is defined in (\ref{DWenergy})
and the energy of the CK is

\begin{align}
\mathcal{E}_\mathrm{CK}&=\sqrt{\frac{k\pi}{ u}-\frac{h\pi}{2\sqrt{2} u^{3/2}}\ln\frac{k+\sqrt{2u}}{k-\sqrt{2u}}}
\label{E_CK}
\end{align}

As $u\to\infty$ for fixed $h$, this chiral kink energy upper bound goes to 0, $\mathcal{E}_\mathrm{CK}\to u^{-1/4} $, so we expect the chiral kink on a stripe to remain stable for large $u$.

It is worth emphasizing that the diagram for skyrmion stability in Fig.~\ref{fig:hu} should be understood as an estimation for the upper bound range for the stability of skyrmions with CKs. 
At any point inside the range bound by the $h_\mathrm{c}$ and $h_\mathrm{e}$, there is an exponentially localized solution for chiral skyrmion with CKs, but this does not guarantee that such a solution will be stable everywhere inside the shaded region. 
As a final note, we  point out that solutions with indefinite chirality may also exist outside the shaded area, for instance, so-called in-plane skyrmions at very strong easy-plane anisotropy. 
However, according to the arguments provided in Ref.~\onlinecite{Kuchkin_20}, such solutions should be considered as a distinct class of soliton-like solutions composed of coupled vortices and antivortices.
A distinguishing feature of these solutions is that the saturated ferromagnetic state representing the vacuum for such vortex-like solutions has a nonzero in-plane component of magnetization.
The energy of that state is degenerate with respect to the rotation around the plane normal.

\section{Characteristic size of chiral kinks}
\begin{figure}[h]
\centering
\includegraphics[width=8cm]{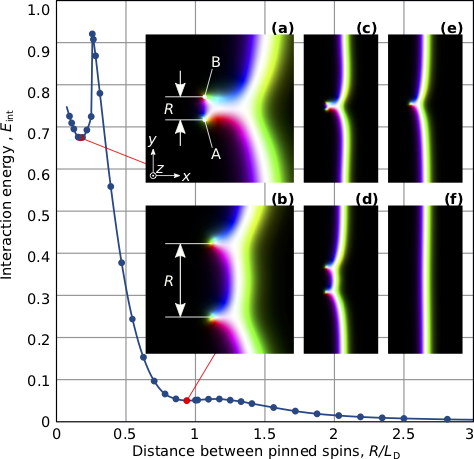}
\caption{The interaction energy between two negative CKs at different distance, $R$. The points A and B indicate the position of fixed spins with $\Theta=\pi/2$ and $\Phi_\mathrm{A}=3\pi/4$ and $\Phi_\mathrm{B}=\pi/4$, respectively.
(a) and (b) are zoomed images of (c) and (d), respectively.
(e) and (f) are isolated stripe with one negative CK and the stripe without kinks, respectively.
The parameters of the system: $h=0.65$, $u=0$, domain size $3L_\mathrm{D}\times6L_\mathrm{D}$, mesh density $\Delta l=128$.
}
\label{fig:interaction}
\end{figure}

Since the typical soliton solutions are exponentially localized, there is not a unique approach to measuring their size, but there are a few conventional approaches to estimating it.
The same is true for CKs which  in addition represent only an element of the spin-texture and cannot be  treated  as isolated objects.
To estimate the characteristic sizes of CKs we suggest an approach based on the analysis of the interaction between them.
Fig.~\ref{fig:interaction} shows the numerically calculated potential energy dependence of two positive CKs located on one side of the isolated stripe. 
The interaction energy, $E_\mathrm{int}$ is defined as follows
\begin{equation}
    E_\mathrm{int}(R)=E_\mathrm{tot}(R)-2E_\mathrm{CK}-E_\mathrm{IS},
\end{equation}
where $E_\mathrm{tot}(R)$ is the total energy of the state composed of two negative CKs, see for instance the  equilibrium states depicted in  Fig.~\ref{fig:interaction}~(a)-(d), 
$E_\mathrm{CK}=E_\mathrm{IS+CK}-E_\mathrm{IS}$ is the self energy of an isolated negative CK on the isolated stripe [Fig.~\ref{fig:interaction}~(e)], and 
$E_\mathrm{IS}$ is the self energy of an isolated stripe without CK [Fig.~\ref{fig:interaction}~(f)]. 
The distance $R$ is the distance between two points A and B with fixed spins:
$$R=|\mathbf{r}_\mathrm{A}-\mathbf{r}_\mathrm{B}|,$$
where $\mathbf{r}_\mathrm{A}=(x_0,-R/2)$ and $\mathbf{r}_\mathrm{B}=(x_0,R/2)$ are the position vectors in two-dimensional plane, $x_0$ may have any arbitrary chosen value.
The two pinned spins at points  $\mathbf{r}_\mathrm{A}$ and $\mathbf{r}_\mathrm{B}$ are lying in plane of the film, $\Theta=\pi/2$, while $\Phi_\mathrm{A}=3\pi/4$ and $\Phi_\mathrm{B}=\pi/4$ (so $\mathbf{n}(\mathbf{r}_\mathrm{A})=(-1/\sqrt{2},1/\sqrt{2},0)$ and $\mathbf{n}(\mathbf{r}_\mathrm{B})=(1/\sqrt{2},1/\sqrt{2},0)$).
An equilibrium position of all other spins is defined via the direct energy minimization scheme.

The interaction energy between two CKs in Fig.~\ref{fig:interaction} has two local minima at $R\approx 0.2L_\mathrm{D}$ and $R\approx 0.9L_\mathrm{D}$ and global minimum at $R\rightarrow \infty$.
The equilibrium configurations corresponding to local minima obtained without spins pinning are shown in Fig.~\ref{fig:interaction} (a) and (b).
The minimum corresponding to the smallest distance $R$ can be thought of as a reasonable estimate for the characteristic size of the CK.

The presence of two local minima indicates the presence of two characteristic scales of inhomogeneities in the system, which is not typical for the majority of  magnetic systems and represents an intriguing feature of chiral magnets.
The latter also allows classifying the different types of solutions according to the inherent scale of inhomogeneities.
For instance, the solutions free of CKs such as spin-spirals, $k\pi$-skyrmions and a variety of skyrmion sacks discussed in Ref.~\onlinecite{Rybakov_19} one may attribute to the class of the solutions with inhomogeneities at the large scale. 
The characteristic scale of inhomogeneities for this class of solutions is of  the order of the  equilibrium pitch of the helical spiral, $L_\mathrm{D}$.
Accordingly, the skyrmions with CKs presented in this paper can be attributed to the class of solutions with magnetic inhomogeneities at small scale -- about an order of magnitude lower than $L_\mathrm{D}$.

An important consequence of the above is that for numerical analysis of functional (\ref{eq:E_totm}) one has to use the finite difference scheme with an appropriate mesh density. 

\section{Skyrmions with negative chiral kinks}

\begin{figure}[h]
\centering
\includegraphics[width=8cm]{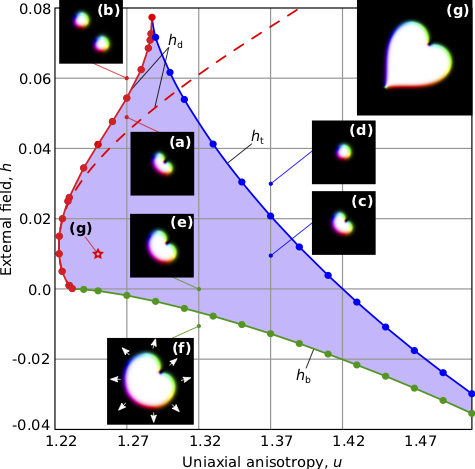}
\caption{$h$-$u$ diagram of stability of the chiral skyrmion with negative CK and $Q=-2$ \textbf{(a),(c),(e)}. Below the blow-up-field, $h_\mathrm{b}$ (green line), the skyrmion starts to expand abruptly (f). 
With increasing field the skyrmion with one negative CK may abruptly split into two $\pi$-skyrmions (b). This skyrmion duplication transition is marked as  $h_\mathrm{d}$ (red solid line). 
Above the critical field, $h_\mathrm{t}$ the skyrmion with $Q=-2$ transforms into a single $\pi$-skyrmion (d).  
The inset (g) illustrate the most exotic and least stable skyrmion among other presented in the paper solutions. The skyrmion in (g) is stable in the narrow range near $h\!=\!0.01$, $u\!=\!1.25$ and contains one positive CK and one negative CK, $Q=-1$. 
The parameters of the system: domain size $5L_\mathrm{D}\times5L_\mathrm{D}$, mesh density $\Delta l=64$.
With increasing $\Delta l$ the critical field $h_\mathrm{d}$ tends to dashed (red) line, while $h_\mathrm{t}$ shifts towards $u\rightarrow \infty$.
}
\label{huQ-2}
\end{figure}

Up to now, we have considered a variety of skyrmions with positive CKs  while for skyrmions with negative CKs we found only a few solutions. The stability range of these solutions is smaller than for skyrmions with positive CKs and requires strong easy-axis anisotropy.
We have estimated the range of existence for the most stable solution with $Q=-2$ [Fig.~\ref{huQ-2}].
The stability region of this solution is bounded by three distinct critical lines, which correspond to three different mechanisms of collapse.
In the inverted magnetic field, $h<0$, the skyrmion blows up at the so-called bursting field, $h_\mathrm{b}$.
On the other hand for $h>0$ the skyrmion with negative CK may either converge to two $\pi$-skyrmions at $h>h_\mathrm{d}$ via  a  duplication mechanism, or may transform into a single $\pi$-skyrmion $h>h_\mathrm{t}$.
For the chosen value of mesh density ($\Delta l = 64$) the two critical fields $h_\mathrm{d}$ and $h_\mathrm{t}$ meet at $u=1.287$.
However, when  increasing the accuracy of the calculations by increasing the mesh density and thereby approaching the continuum limit, the critical field $h_\mathrm{d}$ converges to the values depicted as a dashed (red) line.
Moreover, with increasing $\Delta l$, the whole curve $h_\mathrm{t}$ quickly shifts to the right, towards higher $u$.
For instance, for $\Delta l > 128$, we did not find any evidence for a corresponding field instability for any reasonable $u$.
Therefore, in the continuum limit the stability range for a skyrmion with one negative CK similar to the one shown in  Fig.~\ref{fig:hu} is limited only by two critical fields $h_\mathrm{b}$  and $h_\mathrm{d}$ (dashed red line).

An important aspect of the diagram shown in Fig.~\ref{huQ-2} is that within the stability range for the $Q\!=\!-2$ skyrmion all the solutions for chiral skyrmions with positive CKs presented in this paper are also stable.
A good candidate for a point where all possible solutions may coexist is the point $h=0.01$, $u=1.25$.
In this point and its narrow environment, we succeeded in stabilizing  the most exotic skyrmion solution for $Q=-1$ containing one negative CK and one positive CK [Fig.~\ref{huQ-2}~(g)]. 

\section*{Conclusion}
In this paper, we showed that the structure of 2D chiral skyrmions can be understood and classified in terms of their  constituent domain walls with or without chiral kinks.
Since chiral kinks produce regions of energetically disfavoured chirality which contribute positively to the total energy, one might  expect spin textures containing chiral kinks  to be unstable. Our results show that this expectation is naive, and  that  a wide range of isolated skyrmions containing chiral kinks is in fact stable.

Remarkably, we found that the topology  of domain walls and chiral kinks in our initial configuration  does not change during energy minimisation in a large portion of the phase diagram. 
Thus we were able to exploit the simplicity of solutions at the Bogomol'nyi point to pick configuration with an interesting domain wall and kink structure and obtain metastable configurations with the same structure at other points in the phase diagram by numerical minimization. The combination of our ansatz in terms of a holomorphic function and subsequent numerical minimisation thus allowed us to obtain many previously unknown types of chiral skyrmions.

We estimated the region of stability for skyrmions with chiral kinks as a function of the external magnetic field $h$ and the anisotropy parameter $u$. In particular, we showed that for any fixed $u>-0.5$ there is a critical magnetic field, $h_\mathrm{c}$, above which these solutions collapse. On the other hand, for any $0<h<h_\mathrm{c}$ one can find a skyrmion with a chiral kink at any sufficiently big anisotropy $u\gg0$. 

In  this paper, we  also took the first steps in studying the structure and interactions of chiral kinks. For chiral kinks on a single straight stripe we analysed the profile of a single kink and studied  the interactions of  two  kinks. In a forthcoming paper~\cite{letter},  we also study chiral kinks on  curved domain walls, and  look at the interplay between the   curvature of the  domain wall and the  chiral kinks on it.

The use of holomorphic data for initial configurations provides a versatile and powerful tool but it has also certain limitations. For example, the presented approach does not provide an ansatz for skyrmion bags of negative topological charge. Since the skyrmion bag is a closed $2\pi$-domain wall that contains the skyrmions in it, the repulsion of the skyrmions on the inside  must be balanced by the tension (tendency to shrink) of the domain wall. 
While closed circular  $2\pi$-domain walls are possible at the Bogomol'nyi point, they have an arbitrary size which results in zero tension that cannot balance the repulsive forces of $Q=-1$ skyrmions.

We end with a brief outlook on the interaction of several localised  chiral skyrmions. 
Based on  our preliminary numerical results we conjecture that configurations with chiral kinks on an outer domain wall will generally attract while chiral skyrmions without kinks on their outer domain wall to repel. It would clearly be interesting to test this conjecture with further numerical analysis or to prove it analytically. We leave this as a  challenge for future work.

Experimental observations of  domain walls with chiral kinks  were recently reported in Pt/Co/Ni/Ir multilayers~\cite{Li_2020}.  One may therefore expect that  tuning   the parameters  of  such a system will also allow  experimental observations of some of the skyrmions presented in this work.

\begin{center}
    {\footnotesize\bf{ACKNOWLEDGMENTS}}
\end{center}
The authors acknowledge financial support from the European Research Council (ERC) under the European Union's Horizon 2020 research and innovation program (Grant No. 856538, project "3D MAGiC"),  from Deutsche For\-schungs\-gemeinschaft (DFG) through SPP 2137 "Skyrmionics" (Projects KI 2078/1-1 ,  BL 444/16).

\appendix
\section{Winding numbers of the domain wall}\label{DWWN}
To prove the formula  \eqref{QBloch} for the degree of a configuration $\mathbf{n}$, we note that, in terms of the vector field
\begin{align}
    \mathbf{g}= \sin\Theta\, \nabla\Theta\times \nabla \Phi,
\label{gdef}    
\end{align}
the integrand of \eqref{Qint} can be written as 
\begin{align}
  \mathbf{n}\cdot \left(\partial_\mathrm{x}\mathbf{n}\times\partial_\mathrm{y}\mathbf{n}\right) = \mathbf{e}_\mathrm{z}\cdot \mathbf{g} ,
  \label{gyro}
\end{align}
and so  the degree may also be viewed as the total flux through  the plane of the emergent field $\mathbf{g}$. 
This field  is not generally  globally a curl (otherwise its flux would be zero) but it  can be written as the curl of  Dirac's monopole vector potentials $\mathbf{a}_+$ and $\mathbf{a}_-$ in the  positive and negative domains, respectively:
\begin{align}
  \mathbf{g} = \nabla \times \mathbf{a}_\pm, \;\; \mathbf{a}_\pm = (\pm 1 - \cos\Theta)\nabla \Phi.  
\end{align}
Splitting the integral \eqref{Qint},   with the integrand written according to  \eqref{gyro},  as a sum of integrals over positive and negative  domains, applying Stokes's theorem in each domain and  keeping track of the orientation of bounding Bloch walls then yields \eqref{QBloch}.

We define the kink field as the azimuthal angle $\Phi(s)$ relative to the angle  $\varphi(s)$ which the tangent vector to the curve makes with the $x$-axis, i.e. $\Psi(s) =\Phi(s) -\varphi(s)$. 

The signed curvature of a domain wall is given by
\begin{align}
\frac{\mathrm{d} \varphi}{\mathrm{d}  s} = \kappa(s).
\label{curv}
\end{align}
we use  \eqref{curv} to deduce
\begin{align}
\label{PhiPsi}
\Psi'(s) = \Phi'(s) -\kappa(s).
\end{align} 

Defining  the winding number of the kink field as 
\begin{align}
N_{\text{\tiny kink}}=\frac{1}{2\pi} \int_C \nabla \Psi \cdot \mathrm{d} \mathbf{r},
\label{Ndef}
\end{align}
where we again assume our chosen orientation of $C$, 
we integrate \eqref{PhiPsi}  to deduce 
\begin{align}
N_{\text{\tiny kink}} = w(C)-\iota(C),
\end{align}
where $\iota(C)$ is the winding number of the wall. For simple and closed  curves $C$,   $\iota(C) =\pm 1$  by Hopf's Umlaufsatz \cite{Klingenberg}. Specifically, $\iota(C)=1$ if the orientation of $C$ agrees with its geometrical orientation and $\iota(C)=-1$ otherwise.

\section{ Modification of the chiral magnet energy by a boundary term}

For most  values of the phase diagram parameters $(h,u)$, skyrmion solutions are exponentially localised. In this  appendix we  explain why the standard expression \eqref{eq:E_totm} for the energy  of chiral magnets should be modified along  a  critical  line in the phase diagram where solutions are only localised according to a  power law.  This line includes the Bogomol'nyi point, and the modification  at that point  was addressed in Ref.~\onlinecite{Barton-Singer_20} and more generally in Ref.~\onlinecite{Schroers_20}.  The modification was first   introduced for analytical reasons in  an earlier paper~\cite{Melcher}. Here we illustrate  the consequences of the  modification concretely for axially symmetric configurations.

We consider a  family  of  energy  expressions of the form  \eqref{eq:E_totm}, depending on a parameter $\mu$:
\begin{align}
\mathcal{E}(\mathbf{n})\!=\! \int\!\left(\frac{1}{2}\left(\nabla\mathbf{n}\right)^{2}\!+\!2\pi \mathbf{n\cdot}\nabla\!\times\!\mathbf{n}\!+\!\frac{\mu^2}{2}(1\!-\!n_\mathrm{z})^2 \right)\mathrm{d}x\mathrm{d}y.
\label{Bogoline}
\end{align}
In the phase diagram  parametrised by  $h$ and $u$, this family constitutes a line along which 
the potential changes from having a unique minimum, favouring a ferromagnetic phase, to having a circle of minima, favouring a symmetry-breaking tilted ferromagnetic phase. The value $\mu=2\pi$
 defines the Bogomol'nyi point. 

With this energy one can find~\cite{Melcher, Barton-Singer_20} an exact hedgehog solution (${\Phi\!=\!\phi + \pi/2}$) to the Euler-Lagrange equations with profile
\begin{equation}
\label{hedgehog}
\Theta(r) = 2 \arctan\left(\frac{4\pi}{\mu^2  r}\right).
\end{equation}
At the Bogomol'nyi point $\mu = 2\pi$, this reduces to the solution \eqref{BG_solution} with vanishing holomorphic part $f(\zeta)=0$.

However, although \eqref{hedgehog} solves the Euler-Lagrange equations it is not a stationary point of the energy with respect to scaling of the solution. This can be seen by the usual Derrick argument. Writing $\mathcal{H}, \mathcal{W_D}$ and $\mathcal{U}$ for the  integrated exchange energy, DMI energy and potential energy, the total energy of a rescaled configuration  $\mathbf{n}_\lambda(\mathbf{r}) =\mathbf{n}(\lambda \mathbf{r}) $ is
\begin{equation}
\mathcal{E}(\mathbf{n}_\lambda) = \mathcal{H}(\mathbf{n}) + \frac{1}{\lambda}\mathcal{W_D}(\mathbf{n}) +  \frac{1}{\lambda^2}\mathcal{U} (\mathbf{n}).
\end{equation}
If we want our configuration to be a stationary point with respect to this scaling, we require:
\begin{equation}
\frac{\partial}{\partial \lambda}\mathcal{E}(\mathbf{n}_\lambda)|_{\lambda=1} = -\frac{1}{\lambda^2}\mathcal{W_D}(\mathbf{n}) -  \frac{2}{\lambda^3}\mathcal{U}(\mathbf{n}) =0,
\end{equation}
so the contribution to the energy from the DMI must be $-2$ times the contribution from the potential. 
These arguments are common for a general form of the potential~\cite{Bogdanov_95}.
However, if we evaluate the various terms for  our exact hedgehog solution, we instead find  $\mathcal{W_D} = -\mathcal{U}$, showing that  it cannot be a  stationary point of the energy under scaling.  So, even though the hedgehog  \eqref{hedgehog}  solves the  Euler-Lagrange equations, it  cannot be  a stable minimum of the energy.

This problem arises because the hedgehog configuration \eqref{hedgehog} falls off like $1/r$ at infinity, and  therefore re-scaling is a variation which also falls of like $1/r$. While the solution  \eqref{hedgehog} is a stationary point of the energy  with respect to variation with vanish rapidly at infinity, it is not a stationary point with respect to variations which decay like $1/r$. One  can fix this problem by modifying the energy functional by subtracting the total vorticity \cite{Barton-Singer_20}, \cite{Schroers_20}
\begin{align}
\tilde{\mathcal{E}}(\mathbf{n})=\mathcal{E}(\mathbf{n}) -2\pi\int (\partial_\mathrm{x} n_\mathrm{y} - \partial_\mathrm{y} n_\mathrm{x}) =\nonumber\\
\int\left( \frac{1}{2}(\nabla \mathbf{n})^2 + 2\pi (\mathbf{n}-\mathbf{e}_\mathrm{z})\cdot (\nabla \times \mathbf{n})+\right.\\
\left.4\pi^{2}\frac{\mu^2}{2}(1-n_\mathrm{z})^2  \right)\mathrm{d}x\mathrm{d}y\nonumber 
\end{align}

The modified energy functional only differs from the original energy functional for configuration which decay like $1/r$ at infinity. For faster decays it does not change the total energy.

With this correction, we solve the problem above: now any finite-energy solution of the Euler-Lagrange equations is a stationary point of the energy. And in particular, the exact solutions we have,  for arbitrary $\mu$ and in particular  at the Bogomol'nyi point, are minima of the energy with respect to any variation. This means that the energy of  a configuration is simply bounded below by its degree. At the Bogomol'nyi point, one finds that the  skyrmion ($Q=- 1 $ and anti-skyrmion ($Q=1$)    solutions   have energy $-4\pi$ and $+4\pi$, while without the correction they were degenerate in energy. This is the ordering we expect physically, and so provides another justification for working with the modified energy expresseion. Finally, with this correction the energy of the  domain wall  solution \eqref{bogowall}  is well-defined and equal to zero at the Bogomol'nyi point.

 \section{Chiral kink ansatz}\label{CKansatz}

 For describing the profile of the stripe with a kink we use the ansatz functions method and let $\Theta=\Theta(x)$ as in (\ref{bogowall}) and let $\Phi$ depend on $y$ for $x>0$ only.
 Substituting this into energy functional (\ref{eq:E_E0}) and deciding that a single kink energy $\mathcal{E}_\mathrm{CK}^{a} = \mathcal{E}_\mathrm{IS+CK}^{a} - \mathcal{E}_\mathrm{IS}$ we have

 \begin{align}
 \mathcal{E}_\mathrm{CK}^{a} = \int_{-\infty}^{\infty}\left(\frac{I}{2}\left(\frac{\mathrm{d}\Phi}{\mathrm{d}y}\right)^{2}+2\pi^{2}\left(1-\sin\Phi\right)\right)\mathrm{d}y,
\label{E_kink_int}
 \end{align}
 where $I=\int_{0}^{\infty}\sin^{2}\Theta \mathrm{d}x$. $I$ can be written in the form

 \begin{align}
 I=\frac{k}{2\pi u}-\frac{h}{4\sqrt{2}\pi u^{3/2}}\ln\frac{k+\sqrt{2u}}{k-\sqrt{2u}},
 \label{I_int}
 \end{align}

\begin{figure}[h]
 \centering
 \includegraphics[width=8cm]{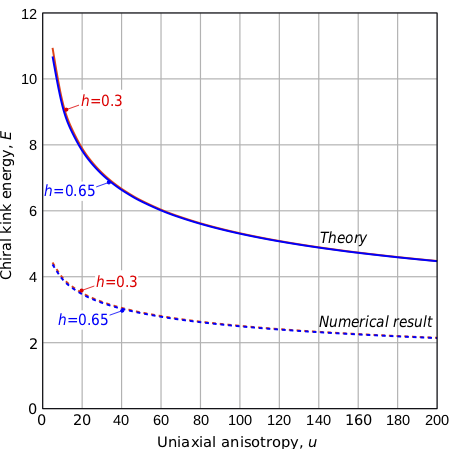}
 \caption{Energy dependency of a kink energy $\mathcal{E}_\mathrm{CK} = \mathcal{E}_\mathrm{IS+CK} - \mathcal{E}_\mathrm{IS}$ as a function of anisotropy, $u$, for two fixed values of magnetic fields  $h=0.3$ and $h=0.65$. Solid lines correspond to $\mathcal{E}_\mathrm{CK}^{a}$, dashed lines are result of numerical minimization. For simulations mesh densities $\Delta l=256,512$ are used.}
 \label{kink_energy}
 \end{figure}

 Treating the profile of $\Theta(x)$ as fixed, we can minimize the energy with respect to $\Phi(y)$ to get $\Phi(y)=4\arctan(e^{m y})+\pi/2$, where $m=\sqrt{\frac{2}{I}}\pi$ 
 and $\mathcal{E}_\mathrm{CK}^{a} = 8\sqrt{2I}\pi$. Note that in this calculation, the contribution for the DMI has acted like a potential for the function $\Phi$ along the domain wall. Extending this observation to allow for a wall that changes shape in response to $\Phi$ is the subject of a future paper \cite{letter}. This value for the energy is shown on Fig.\ref{kink_energy}.

We consider  the limiting case $u\rightarrow\infty$ in formula (\ref{I_int}). In this case we obtain $I \rightarrow \frac{1}{\pi\sqrt{2u}}\left(1+\frac{h}{2}\frac{\ln u}{u}\right)$ and the corresponding kink energy decreases as $\mathcal{E}_\mathrm{CK}^{a} \rightarrow \frac{8\sqrt{2\pi}}{(2u)^{1/4}}\left(1+\frac{h}{4}\frac{\ln u}{u}\right)$. The analytic expression for the  exact kink energy (see numerical curve on Fig.\ref{kink_energy}) is  unknown but the inequality $\mathcal{E}_\mathrm{CK} \leq \mathcal{E}_\mathrm{CK}^{a}$ holds for any ansatz functions. Therefore the exact kink energy is bounded: $\mathcal{E}_\mathrm{CK} \leq\frac{8\sqrt{2\pi}}{(2u)^{1/4}} + \mathop{O}\left(\frac{\ln u}{u^{5/4}}\right)$ .
The fact that the energy of the ansatz goes to $0$ at high $u$ suggests that this is a good approximation to the exact kink solution at high $u$. This fits with the observation that at high $u$, the wall is `stiff' and does not bend much in response to the kink.

\end{document}